\documentstyle [12pt,axodraw] {article}

\catcode`@=12
\def\slashchar#1{\setbox0=\hbox{$#1$}           
   \dimen0=\wd0                                 
   \setbox1=\hbox{/} \dimen1=\wd1               
   \ifdim\dimen0>\dimen1                        
      \rlap{\hbox to \dimen0{\hfil/\hfil}}      
      #1                                        
   \else                                        
      \rlap{\hbox to \dimen1{\hfil$#1$\hfil}}   
      /                                         
   \fi}
\def\pslash{\slashchar{p}}

\def\H12{H_{12}}
\def\H12t{\widetilde{H_{12}}}

\begin{document}
\thispagestyle{empty}
\begin{flushright} August 1999\\
hep-ph/9908417
\end{flushright}
\vspace{0.5in}
\begin{center}
{\Large \bf Leptogenesis with 
Heavy Majorana Neutrinos Reexamined\\}
\vspace{0.7in}
{\bf Raghavan Rangarajan$^*$ and Hiranmaya Mishra$^\dagger$ 
\\}
\vspace{0.2in}
{ \sl Theoretical Physics Division\\
Physical Research Laboratory, Ahmedabad 380 009, India\\}
\vspace{0.1in}
\end{center}
\begin{abstract}

The mass term for Majorana neutrinos explicitly violates lepton number.
Several authors have used this fact to create a lepton asymmetry
in the universe by considering CP violating effects in
the one loop self-energy correction for the decaying
heavy Majorana neutrino.  We compare and comment on the different 
approaches used to calculate the lepton asymmetry including those using
an effective Hamiltonian and resummed propagators.
We also recalculate the asymmetry in the small mass difference limit.

\end{abstract}

\vspace{1in}

* E-mail: raghavan@prl.ernet.in

$\dagger$ E-mail: hm@prl.ernet.in

\newpage
\baselineskip 18pt

After Sakharov's seminal paper on the conditions for obtaining a matter
asymmetric universe several authors have attempted to realise these
conditions in the context of different models\cite{reviews}. 
One approach is to create a
lepton asymmetric universe by the asymmetric decay of heavy scalars or Majorana
neutrinos and to convert the lepton asymmetry into a baryon asymmetry by B+L
violating sphaleron processes. In the original papers on leptogenesis CP
violation was realised by the interference between the tree level diagram
and the one loop vertex correction\cite{leptov}. 
Later it was realised that
there was an additional contribution from the one loop self-energy
correction to the external heavy scalar or
neutrino\cite{IKS,LS}.

The self-energy contribution has been calculated differently by different
authors. Refs.\cite{LS,CRV96} obtain the asymmetry by considering the
interference of the tree level diagram for decay and the diagram
containing the one loop self-energy correction.
However
this approach breaks down in the degenerate heavy particle mass limit.
In
refs.\cite{FPS,FPSW,ma,SV,CR97} 
the authors obtain the one loop effective Hamiltonian, or more
precisely, the one loop effective mass matrix for the heavy species.
Because particles in the self-energy loop go on shell the effective mass
matrix is non-Hermitian. This, coupled with CP violation,
results in different evolution of the scalars and
their antiparticles, or of the right-handed and left-handed neutrinos,
and ultimately gives rise to an asymmetry when the
heavy particles decay. In refs.\cite{FPSW,CR97} 
the authors find an enhancement
in the degenerate scalar or neutrino mass limit\cite{KRS}. 
The authors of
refs.\cite{FPS,FPSW,ma,SV} and \cite{CR97} 
differ in their use of the effective Hamiltonian.
In this
paper we compare and comment on both approaches
and on the approach in 
refs.\cite{BP,pilaftsis2} where the authors use 
resummed propagators to evaluate cross sections
and decay rates.
We shall discuss below models involving decays of heavy Majorana
neutrinos.  However our comments also apply to models with heavy scalars.
In refs.\cite{ma,CR97} the authors consider models with charged scalar
fields; to facilitate discussion 
we have modified their definitions to suit the Majorana fields
$N_a$.  

In section 1 we discuss some subtleties associated with a non-Hermitian
Hamiltonian, particularly with respect to the normalisation
of eigenstates of the Hamiltonian.  In section 2 we compare the approaches of refs.\cite{FPSW}
and \cite{CR97} both of whom use the effective Hamiltonian.
In section 3 we discuss some field theoretic concerns regarding
the definition of the decay
amplitude for eigenkets of a non-Hermitian Hamiltonian 
and the existence of a Fock space for mixed
fields.  In section 4 we compare the 
matrices used to diagonalise the effective Hamiltonian or projections
of the resummed 
propagator in
refs.\cite{FPSW,SV,BP}.  We also comment on the approach
of ref.\cite{pilaftsis2}.  We re-evaluate the asymmetry in the small 
mass difference limit.  In section 5 we summarize our results.

\section{Diagonalisation of the effective mass matrix}

Let us start with the lagrangian
\begin{eqnarray}
{\cal L}_{int} & = & \sum_{i} M_i [\overline{(N_{Ri})^c} N_{Ri} +
   \overline{N_{Ri}} (N_{Ri})^c] \nonumber \\
 & & + \sum_{\alpha, i} \, h^\ast_{\alpha i} \, \overline{N_{Ri}}
 \, \phi^{\dagger} \,
     \ell_{L}^{\alpha} + \sum_{\alpha, i} h_{\alpha i} \,
      \overline{\ell_{L}^{\alpha}} \phi \, N_{Ri} \label{eq:lag}\\
 & & + \sum_{\alpha, i} \, h^\ast_{\alpha i} \, \overline{(\ell_{L}
^{\alpha}})^c \, \phi^\dagger \, (N_{Ri})^c + \sum_{\alpha, i}
 h_{\alpha i} \overline{(N_{Ri})^c} \,
       \phi \, (\ell_{L}^{\alpha})^c  \nonumber
\end{eqnarray}
as in ref.\cite{FPSW}. $N_{iR}^c=(N_{iR})^c$ and $N^c=C\bar N^T$. 
$l_L^\alpha$  are light leptons and $\phi$ is a scalar field.  For
concreteness the $l_L^\alpha$ can be the left-handed lepton doublet
of the Standard Model, $\phi$ can be the Standard Model Higgs doublet.
We supress the SU(2) indices carried by the 
lepton and the Higgs fields which are contracted
by the $\epsilon_{ab}$ tensor.
$\alpha$ and $i$ are generation indices and we assume one heavy Majorana field
per generation of light leptons.  
For simplicity we shall work with two 
generations.  The lagrangian may also be rewritten in 
terms of $\eta_i=N_{iR}+N_{iR}^c$.  In what follows we shall replace
$N_{iR}^c$ by $N_{iL}$.

Since the right handed neutrinos $N_{iR}$ decay to leptons while
the left-handed neutrinos $N_{iL}$ decay to antileptons,
the lepton number violation is in the mass
term for the heavy neutrinos. Note that the last two lines in the lagrangian
above are equivalent as 
$\bar{\chi _1}\chi_2=\overline{\chi _2^c}\chi _1^c$.
One loop effects give rise to mixing between heavy neutrino flavours as
shown in fig. 1. 
\begin{figure}
\mbox{}
\vskip 4.25in\relax\noindent\hskip .2in\relax
\includegraphics{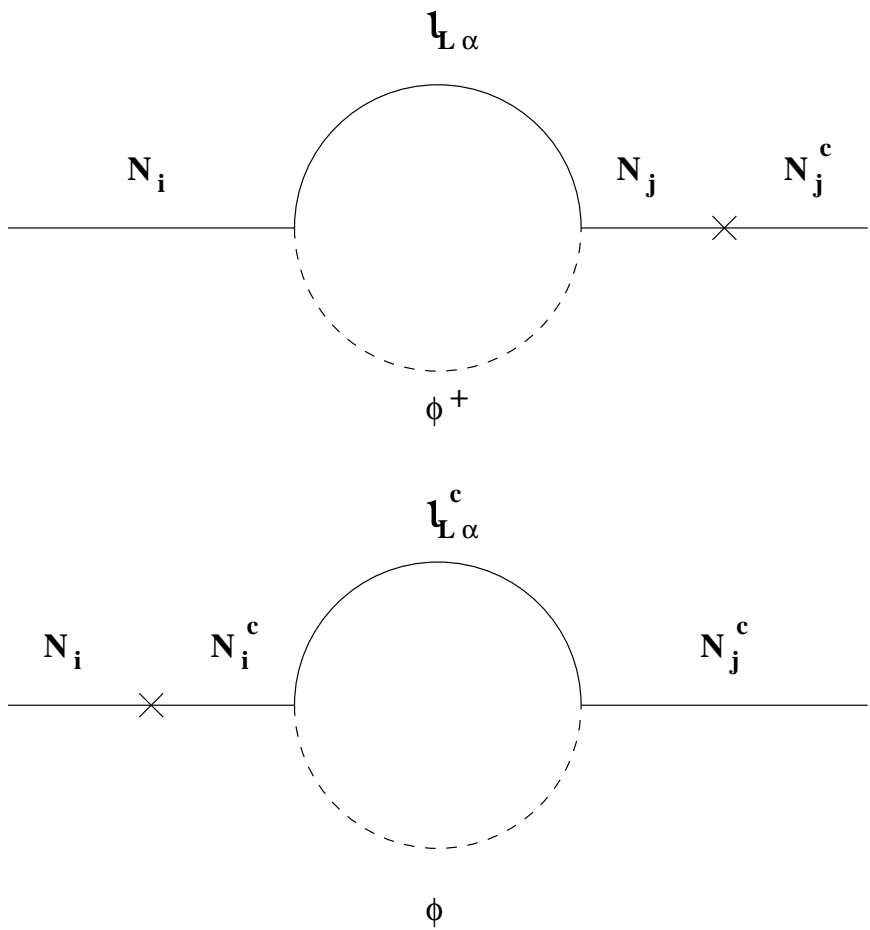}
\caption{ One loop contributions to the mass matrix}
\end{figure}
The one loop effective mass term due to the diagrams in
fig. 1 is given by
\begin{equation}
\left(
\bar N_{1L}\,\bar N_{2L}\,\bar N_{1R}\,\bar N_{2R}\,
\right)
 \left( \begin{array}{cccccc}
                  0 & 0 & M_1+H_{11} & H_{12} \\
                  0 & 0 & H_{12} & M_2+H_{22} \\
                  M_1+H_{11} & \widetilde{H}_{12} & 0 & 0 \\
                  \widetilde{H}_{12} & M_2+H_{22} & 0 & 0 \\
              \end{array}    \right)
\left( \begin{array} {c} N_{1L} \\ N_{2L} \\ N_{1R} \\ N_{2R}\\
\end{array} \right)
\label{eq:Meff}\end{equation}

Ignoring the masses of the light leptons and of the scalar field,
the expressions for $H_{ij}$and $\widetilde H_{ij}$
are given in ref.\cite{FPSW} as
\begin{equation}
H_{ij} = H_{ji} =
\left[ M_i \sum_\alpha h_{\alpha i}^\ast h_{\alpha j} +
 M_j \sum_\alpha h_{\alpha i} h_{\alpha j}^\ast \right] (g_{\alpha ij}^{dis}
   - \frac{i}{2} g_{\alpha ij}^{abs})
\end{equation}
\begin{equation}
\widetilde{H}_{ij} = \widetilde{H}_{ji} =
\left[M_i
                               \sum_{\alpha} h_{\alpha i} h_{\alpha j}^\ast
     + M_j \sum_{\alpha} h_{\alpha i}^\ast h_{\alpha j} \right]
     (g_{\alpha ij}^{dis} - \frac{i}{2} g_{\alpha ij}^{abs})
\end{equation}
and
\begin{equation}
H_{ii}^{(1)} = \widetilde{H}_{ii}^{(1)} =
\left[2 M_i \sum_{\alpha}  h_{\alpha i} h_{\alpha i}^\ast  \right]
     (g_{\alpha ii}^{dis} - \frac{i}{2} g_{\alpha ii}^{abs})
\end{equation}
where $g_{\alpha ij}^{abs} =  {1 \over 16 \pi }$.

The effective mass matrix $M_{eff}=M_0+H$ in eq.~(\ref{eq:Meff}) can 
be written as $M-{i\over2}\Gamma$
where the dispersive part of $H$ (i.e. terms proportional to 
$g_{\alpha ij}^{dis}$) is included in $M$ and the absorptive part 
(proportional to $g_{\alpha ij}^{abs}$) is
included in $\Gamma $. The absorptive part reflects the fact that the
particles in the loop have gone on shell. Both $M$ and $\Gamma $ are
Hermitian but not $M_{eff}$. This is analogous to the formulation used to
obtain an effective mass matrix for neutral kaons\cite{paschos}. 
The one loop corrections
in $M$ 
can be absorbed away by renormalisation. 
(The off-diagonal elements of $M$ are renormalised by non-diagonal elements
of the wave function counterterms\cite{kniehlpilaftsis}.)
However, there are no absorptive counterterms\cite{kniehlpilaftsis}, 
nor are they
needed for the absorptive parts are finite. 
It will be the mixing via the absorptive terms that will be important for
creating the lepton asymmetry.

Since the effective mass matrix is non-Hermitian it can
not be diagonalised by a unitary matrix, i.e. $VM_{eff}V^{-1}=M_D$, where $%
M_D$ is diagonal, but $V^{-1}\neq V^{\dagger }$. 
The columns of $V^{-1}$ are the eigenkets of $M_{eff}$ and the
rows of $V$ are the suitably normalised eigenbras of $M_{eff}$, i.e., $%
xM_{eff}=\lambda x$. (The subtleties of normalisation are discussed below.)
If $M_{eff}$ were Hermitian the eigenbras of $M_{eff}$
would be the Hermitian conjugates of the eigenkets and $V^{-1}$ would equal $%
V^{\dagger }$. This is not true in our case, though the bra-eigenvalues and
ket-eigenvalues are the same.

Therefore we rewrite the mass term of eq.~(\ref{eq:Meff}) as 
\begin{equation}
\bar N M_{eff} N=\bar{\tilde\psi} M_D\psi=\bar\psi M^\prime\psi
\label{eq:MD}\end{equation}
where $\psi=V N$ and $\bar{\tilde\psi}=\bar N V^{-1}$, and
only $M_D$ is diagonal above.
Besides the choice mentioned above one can diagonalise
$M_{eff}$ by a bi-unitary transformation, i.e, $U^\dagger M_{eff} V=M_D$, where 
$U$ and $V$ are unitary.
However in this case the kinetic energy term will not be diagonal in the new
$(\bar{\tilde\psi},\psi)$ basis.

If one can expand the $\psi,\tilde\psi$ and $N$ fields in terms of
annihilation and creation operators
(later we comment on this assumption),
then $|\tilde\psi_c\rangle=|N_a\rangle V_{ac}^{-1}$, where
$|\tilde\psi_c\rangle$ represents a particle created by
the action of the creation operator in $\bar {\tilde\psi_c}(x)=\bar N_a(x) V_{ac}^{-1}$ on the vacuum
and $|N_a\rangle$ represents a particle created by the creation operator
in 
$\bar N_a(x)$ acting on the vacuum.  
Note that while $\sum_a|N_a\rangle\langle N_a|=1$,
$\sum_a |\psi_a\rangle\langle\psi_a| \ne 1$
where 
$\langle\psi_a|=|\psi_a\rangle^\dagger$
\footnote{%
In ref.\cite{SV} the authors discuss the fact that because of the
non-Hermitian nature of the effective mass matrix $\psi\ne\psi^c$.
They state that $\psi$ still represents a Majorana field and hence
we sum only over particle states.
In models 
involving charged scalars one certainly needs to consider both particle and
antiparticle states.  In that case the sum over all states will include
particle and antiparticle states.}.
Instead $\sum_a |\tilde \psi_a\rangle\langle\psi_a|
=1$
and 
$\langle\psi_a|\tilde\psi_a\rangle=1$
where $\langle\psi_a|$ represents a particle created
by the action of the destruction operator in $\psi(x)$
acting to the
left on the vacuum%
\footnote{
The normalisation of $\psi$ states has also been discussed elsewhere, 
as for example,
in ref.\cite{LS2} and references thereof.}.

\section{Obtaining the lepton asymmetry}

Refs.\cite{FPS,FPSW,ma,SV} 
and
\cite{CR97} differ in their choice of the initial state relevant for
the calculation of the lepton asymmetry.  Let $\epsilon$ be a measure
of CP violation due to the self-energy correction.  
In refs.\cite{FPS,FPSW,ma,SV}
the authors define%
\begin{equation}
\epsilon_{\psi_a}= \Biggl[{\Gamma_{\tilde\psi_a\rightarrow l} - 
\Gamma_{\tilde\psi_a\rightarrow l^*}\over
\Gamma_{\tilde\psi_a\rightarrow l} +
\Gamma_{\tilde\psi_a\rightarrow l^*}} \Biggr]
\label{eq:epspsi}\end{equation}
while the definition of the asymmetry parameter in ref.\cite{CR97} is
\begin{equation}
\epsilon_{N_a}= \Biggl[{\Gamma_{N_a\rightarrow l} -
\Gamma_{N_a\rightarrow l^*}\over
\Gamma_{N_a\rightarrow l} +
\Gamma_{N_a\rightarrow l^*}} \Biggr]
\label{eq:epsN}\end{equation}
where, as stated earlier, the definitions of the asymmetry parameter
in refs.\cite{ma,CR97} have been modified
for a scenario involving Majorana neutrinos.  Below we compare the approaches of
refs.\cite{FPSW} and \cite{CR97}.

Before we proceed
let us reiterate here the mechanism of the creation of the lepton
asymmetry.  Starting with two relativistic neutrino species
the second (heavier) species becomes non-relativistic at $t_{nr2}$.  After
$t_{nr2}$ the 
physical states are the eigenstates of an effective Hamiltonian for one 
relativistic neutrino species with a Lorentz suppressed mass
and one non-relativistic neutrino species.  They
can be represented by $N_1$ and $\psi_2^\prime$ fields, where $\psi_2^\prime$ is a linear
combination of the $N_a$ (but is not $\psi_2$).  We assume the heavier
$\psi_2^\prime$ decay while the $N_1$ neutrinos are in equilibrium.
At time $t_{eq}$ the lighter
neutrinos $N_1$ go out of equilibrium and later at $t_{nr}$ 
they become non-relativistic, i.e., we assume
$K=M_1/T_{eq}\ll 1$, where $T_{eq}$ is the temperature at $t_{eq}$.  
Thus at $t_{nr}$ one will have 
a universe with only $N_1$ neutrinos with equal numbers of right-handed
and left-handed neutrinos, $N_{1R}$ and $N_{1L}$.  
After $t_{nr}$ 
the effective 
Hamiltonian is as in eq.~(\ref{eq:Meff})
and the $\psi_a$ are the
eigenstates of the effective Hamiltonian.  The state of the universe
with equal numbers of $N_{1R}$ and $N_{1L}$ can now be described as 
a linear combination of the $\psi_a$.  
More precisely, 
a state $|N_{1R}\rangle$ (or $|N_{1L}\rangle$)
at $t_{nr}$ can be rewritten 
in terms of $|\tilde\psi_a\rangle$.
As in the standard neutrino flavour
oscillation scenario, the right- and left-handed content of an originally
$|N_{1R}\rangle$ or $|N_{1L}\rangle$ state 
will now start to oscillate.  One can see from eq.~(\ref{eq:Meff})
that the amplitude
for $|N_{1R}\rangle\rightarrow |N_{2L}\rangle$ is proportional to $H_{12}$ while
the amplitude for $|N_{1L}\rangle\rightarrow |N_{2R}\rangle$ is proportional to 
$\widetilde H_{12}$.  (The amplitudes for $|N_{1R,L}\rangle\rightarrow |N_{1L,R}\rangle$
are equal and proportional to $M_1+H_{11}$.)  Because of CP violation
$H_{12}\ne \widetilde H_{12}$.  Therefore, though one starts
with an equal number of $N_{1R}$ and $N_{1L}$, more $N_{1L}$ may transform
into $N_{2R}$ than $N_{1R}$ into $N_{2L}$ and one ends up with an universe
with an
unequal content of $N_{1R}$ plus $N_{2R}$ and $N_{1L}$ plus $N_{2L}$.
When the 
neutrinos subsequently decay one
obtains a lepton number asymmetry since right-handed neutrinos decay
to leptons and left-handed neutrinos decay to anti-leptons.  (We ignore
here asymmetry due to the vertex correction.)

The formalism of ref.\cite{CR97} starts with a universe
with equal numbers of $N_{1R}$ and $N_{1L}$ at $t_{nr}$, or, equivalently, with states $|N_{1R}\rangle$
and $|N_{1L}\rangle$
with equal probabilities of existing at $t_{nr}$.  
The asymmetry obtained reflects
the oscillations in
the right-left $N$ content that takes place prior to decay.
On the other hand, the final asymmetry is obtained in ref.\cite{FPSW} by
considering only $|\tilde\psi _1\rangle$.  
One may argue that typically one does consider the asymmetry generated by only
the lightest neutrino. But what that entails in this scenario is that the
heavier $\psi_2^\prime$ neutrino 
decays while the interactions of the lighter 
$N_1$ are still in equilibrium, leading to the erasure of any asymmetry generated
in the decay of $\psi_2^\prime$. Subsequently the lightest neutrino $N_1$ goes out
of equilibrium, becomes
non-relativistic 
and is re-expressed as a linear
combination of both $|\tilde\psi _a\rangle$ states.  Thus
one has to consider the lightest
$N_a$ species and not the lightest $\psi_a$ species as the initial state at 
$t_{nr}$.
Nevertheless, since $|\tilde\psi_1\rangle$ is an asymmetric mixture of
$|N_{aR}\rangle$ and $|N_{aL}\rangle$ its decay gives a lepton asymmetry.

Ref.\cite{LS2} also argues against starting with the
$\psi_a$ states as these are ``propagation" states which the authors claim
are not physical.
However our observations are based on a cosmological argument rather than a
field theoretic argument.  (We shall also state the field theoretic
argument later.)   The importance of the initial state is also discussed in
ref.\cite{CR97}.
Our arguments above do not apply to scenarios where $K\gg1$, i.e., when the
neutrinos are non-relativistic when they go out of equilibrium.  
In such cases, from a cosmological standpoint,
$\epsilon_{\psi_a}$ will be the relevant asymmetry parameter.

In the limit of large mass splittings, i.e., $|M_1-M_2|\gg |H_{ij}|$,
the asymmetry obtained in ref.\cite{CR97} is similar
to that obtained in ref.\cite{FPSW}.
We find that 
the approximations needed to obtain this result, i.e.,
ignoring terms proportional to 
$W_1^*W_2$, $W_2^*W_1$ and $|W_2|^2$ in eq.~(17) of ref.\cite{CR97} 
and 
assuming a small mixing angle 
is
equivalent to ignoring the asymmetry from the interference between $\psi_1$ and $\psi_2$ and from $\psi_2$ decay and to
setting $N_1\approx\psi_1$. 
However in the small mass
splitting case the asymmetry obtained in refs.\cite{FPSW,CR97} are not similar,
as we show later.

Our arguments imply that one
has a mixed state (as opposed to a pure state)
of $|N_{1R}\rangle$ and $|N_{1L}\rangle$ with equal
probabilities at $t_{eq}$.  
Thereafter one evolves the $|N_{1R,L}\rangle$ states independently.
Since the mass term is not significant when
$N_1$ is relativistic there will be no significant mixing till $t_{nr}$.
After $t_{nr}$ both $|N_{1R,L}\rangle$ can be rewritten as pure
states of $|\tilde\psi_1\rangle$ and $|\tilde\psi_2\rangle$ and hence one
gets interference between $|\tilde\psi_1\rangle$ and $|\tilde\psi_2\rangle$ states which is
necessary for
left-right oscillations.  However if one argues that
one instead has a mixed state of $|\tilde\psi_a\rangle$ at $t_{nr}$ then one
would evaluate $\epsilon_{\psi_a}$ as in ref.\cite{FPSW}.  In that case the 
initial number densities of the $\psi_a$ particles that enter into the
solution of the Botzmann equation would be weighted by the probabilities of
obtaining the $\psi_a$ at $t_{nr}$, given by $|\langle \psi_a|N_1\rangle|^2$.
One would then include the contributions of both $\psi_a$ to the total 
asymmetry since the asymmetry generated by the decay of $\psi_2$ will not
be erased as $\psi_1$ interactions are out of equilibrium at $t_{nr}$.

\section{Some field theoretic concerns}

We now raise more fundamental issues with both the approaches using the
effective Hamiltonian.
The asymmetry parameter is defined in 
refs.\cite{FPS,FPSW,ma,SV}
using decay rates for rotated fields $\psi_a$ with complex masses.
It is shown in ref.\cite{enz} (in the context of kaon physics)
that the lack of orthonormality of the $|\tilde\psi_a\rangle$ states raises ambiguities in the
normalisation of states making it difficult to interpret $\langle f|\tilde\psi_a\rangle$
as the probability amplitude for $\psi_a\rightarrow f$.  If one 
normalises $|\tilde\psi_a\rangle$ as
$|\tilde\psi_a\rangle/\sqrt{\langle\psi_a|\tilde\psi_a\rangle}$
then the state is scale dependent.
This is because
$|\tilde \psi_a\rangle\rightarrow c|\tilde \psi_a\rangle$ 
implies $\langle\psi_a|\rightarrow\langle\psi_a|c^{-1}$
since $|\tilde\psi\rangle=|N\rangle V^{-1}$
and $\langle\psi|=V\langle N|$.
On the other hand, if one normalises the $\psi$ states as
$|\tilde\psi_a\rangle/\sqrt{\langle\tilde\psi_a|\tilde\psi_a\rangle}$
then $\sum_f |\langle f|\tilde\psi_a\rangle|^2/[{\langle\tilde\psi_a|
\tilde\psi_a\rangle}]
\langle f|f\rangle\ne1$.
This can be better understood by considering the quantum mechanical
example of kaon mixing with $|\tilde\psi_a\rangle$ corresponding to 
$|K_L\rangle$ (or $|K_S\rangle$) and $\langle f|$ corresponding to 
$\langle K_{L,S}|$ and
$\langle \tilde K_{L,S}|$.  A similar argument can be extended to field
theoretic scenarios. 
If $\psi_a$ are intermediate particles, i.e., if one includes a production
amplitude for the $\psi_a$, the above concerns do not apply \cite{enz}.
If one does include a production amplitude for the $\psi_a$ in $\epsilon_{\psi_a}$
it will cancel out in the numerator and denominator leaving behind the 
asymmetry parameter as defined in refs.\cite{FPS,FPSW,ma,SV}.

In ref.\cite{CR97} the asymmetry parameter is defined in terms of decay rates
of the mixed $N_a$ fields.
It has been
pointed out in ref.\cite{kimetal} that in
general there does not exist a Fock space for
mixed fields.  If one assumes that the annihilation operators 
of the $\psi_a$ fields and the creation operators of the $\bar{\tilde\psi_a}$ fields 
obey canonical anticommutation relations
then
the annihilation and creation operators
of the $N_a$ fields satisfy, for example,\cite{kimetal}
\begin{equation}
\{A_i({\bf k},h,t),A_j^\dagger({\bf k^\prime},h^\prime,t)\}
=\delta^3({\bf k}-{\bf k^\prime})\delta_{h h^\prime}
\sum_a  V^{-1}_{ia} V_{aj} {E_a^\prime-h|{\bf k}| \over 2 E_a^\prime} 
\label{eq:FS}
\end{equation}
where $h,h^\prime=\pm1$ represent helicity, 
$E_a^\prime=(|{\bf k}|^2 +m_a^2 -i m_a \Gamma_a)^{1\over2}$, 
the sum is over the mass eigenstates, 
$A_i$ and $A_j^\dagger$ include spinor elements as in ref.\cite{kimetal}
and we have used $N=V^{-1} \psi $ and $\bar N=\bar{\tilde\psi} V$.
These anticommutation relations are non-diagonal in the ${i,j}$ indices
and are hence not canonical.
Therefore if $\bar N=\bar{\tilde\psi} V$, $|N\rangle=|\tilde\psi\rangle V$ 
is not a well defined
statement in general.  
Furthermore, the two cases in which one can define an approximate Fock space
for the $N$ fields, namely, the extreme relativistic limit and almost degenerate
masses with real $V$ \cite{kimetal} do not apply to our problem.

In the limit that all the neutrino species are non-relativistic, as is the
case for $t\geq t_{nr}$, the above commutation relations are diagonal in the
$i,j$ indices.  However, as pointed out in ref.\cite{kimetal}, the annihilation and
creation operators of the $N$ fields are no longer time-independent in this
case.  
But if one starts with an $N_a$ state as the initial state at $t_{nr}$ and
evolves it as a linear combination of $\psi_a$ states then this may not
be a problem since
one requires that the relation $|N\rangle=|\tilde\psi\rangle V$
hold at only one instant, namely, at $t_{nr}$.
Note that in ref.\cite{CR97} the states $|\tilde\psi\rangle$ and $|N\rangle$ are in
co-ordinate space.  The creation and annihilation operators for the mixed
fields in co-ordinate space, namely, $\bar N(x)$ and $N(x)$, do satisfy
diagonal anticommutation relations.

\section{Final comments}

It was pointed out by Veltman that it is inappropriate to have unstable 
particles as asymptotic in or out states\cite{veltman}.  
To avoid this problem different effective
approaches have been adopted to obtain the one loop decay amplitude.
In refs.\cite{FPS,FPSW,ma,SV,CR97} the authors use an one loop effective 
Hamiltonian.  In ref.\cite{BP} the decay amplitude is obtained by splitting
a two particle scattering amplitude with the unstable particle as the 
intermediate particle.  In ref.\cite{pilaftsis2} one obtains the decay amplitude
from an effective LSZ formulation.
We discuss here the differences and similarities in 
the above approaches to obtain the asymmetry.\footnote{%
In section 7.3 of ref.\cite{peskin} a further approach 
to obtain
the decay rate for unstable particles using the optical theorem is given.
In ref.\cite{frereetal} a modified leptogenesis scenario in
which the Majorana neutrinos are virtual intermediate particles is considered 
to get around some of the issues discussed here.
} 

Broadly speaking, one can further divide the different approaches into those
that 
use unrotated fields $N_a$ to define
the initial state, such as refs.\cite{CR97,pilaftsis2}
and those that work with rotated fields to define
the initial state, such as refs.\cite{FPS,FPSW,ma,SV,BP}\footnote{%
It is pointed out in section 5 of ref.\cite{pilaftsis1} that in some cases
diagonalising the effective mass matrix to obtain the $\psi_a$ fields
may not be possible.}.
Though refs.\cite{FPSW,SV,BP}
deal with rotated fields their approaches are somewhat different.  Below we 
first compare the rotating matrices for all these approaches in the small and
large mass difference limits.  We also comment on ref.\cite{pilaftsis2}.
We finally re-evaluate the asymmetry in the small mass difference limit.

In ref.\cite{FPSW} the authors obtain the eigenvalues and eigenvectors
of the effective mass matrix in eq.~(\ref{eq:Meff}).
In ref.\cite{SV} the authors obtain the eigenvalues and eigenkets
of the two off-diagonal
blocks of $M_{eff}$.
In general, the
eigenvalues of the two off-diagonal blocks of the matrix do not equal each
other or the
eigenvalues of the full matrix and the eigenkets of the off-diagonal
blocks describe particles different from those related to the eigenkets
of the full matrix.  

One can show that the eigenvalues of the two off-diagonal blocks are the same
as that of the larger matrix, and the eigenvectors of the larger matrix are
trivial combinations of corresponding eigenvectors of the smaller blocks when
the two off-diagonal blocks are the same.  In the limit of a small mass
difference 
$|M_1-M_2|\ll |H_{12}|$, $H_{12} \approx \widetilde H_{12}$ and we 
find that the diagonalising matrices
are equivalent and one gets the same asymmetry.  For example, the matrix
relating the right handed field projections $N_{aR}$ 
and $\psi_{aR}$ is given by
\begin{equation}
 \left( \begin{array}{cccccc}
                  1 & -1 + {M_1^\prime-M_2^\prime\over 2 H_{12}} \\
                  1-{M_1^\prime-M_2^\prime\over 2H_{12}} & 1  \\
              \end{array}    \right)
\label{eq:VUSres}\end{equation}
for both approaches.  The above matrix is obtained from
$V^{-1}$ of Section 1 and we have
not shown the normalisation.  $M_{i}^\prime=M_i+H_{ii}$.
However in the limit of a large mass difference  
$|M_1+H_{11}-M_2-H_{22}|\gg |H_{12}|$ the relation between 
$N_{aR}$ and $\psi_{aR}$ is given by 
\begin{equation}
 \left( \begin{array}{cccccc}
                  1 & -{M_1^\prime H_{12}+M_2^\prime\H12t\over M_1^{\prime2}-M_2^{\prime2}} \\
                  {M_1^\prime \widetilde H_{12}+M_2^\prime H_{12}\over M_1^{\prime2}-M_2^{\prime2}} & 1  \\
              \end{array}    \right)
\label{eq:VUS4wk}\end{equation}
and
\begin{equation}
 \left( \begin{array}{cccccc}
                  1 & -{H_{12}\over {M_1^\prime-M_2^\prime}} \\
                  {H_{12}\over M_1^\prime-M_2^\prime} & 1  \\
              \end{array}    \right)
\label{eq:VUS2wk}\end{equation}
for refs.\cite{FPSW} and \cite{SV} respectively.
One can see that they are not equivalent.  We find that working with the
2x2 matrices in this limit
gives an asymmetry that is 
approximately 1/2
of the asymmetry obtained using the 4x4 approach.
This can be traced to the absence of the contribution of the term proportional
to $M_1^\prime\widetilde
H_{12}$.  Once
again, if $h_{\alpha i}$ are real $H_{12}=\widetilde H_{12}$ and the 
diagonalising matrices of the
two approaches become identical.

%
%
%

In ref.\cite{ma} the authors also consider the eigenvalues of the two blocks
of a 4x4 effective mass matrix. However our above arguments do not apply to
this case as the two blocks are along the diagonal and are transposes of each
other, necessitating that the eigenvalues of the two 2x2 blocks are the
eigenvalues of the larger matrix and that the eigenvalues of each block are
the same, and the eigenvectors of the larger matrix are a trivial 
combination of the eigenvectors of the blocks.  

Ref.\cite{FPSW} diagonalises the mass matrix 
and obtains the relationship between $|\tilde\psi_a\rangle$ and
$|N_a\rangle$ states given by $|\tilde\psi\rangle =
|N\rangle V^{-1}$. 
On the other hand, ref.\cite{BP} diagonalises projections of the
resummed propagator, $S_{RR}=\langle N_R \bar N_L\rangle$ and $S_{LL}=\langle N_L \bar N_R\rangle$.  
Using this
to factorise the amplitude
for $l\phi^*\rightarrow \psi \rightarrow l^* \phi$
and $l^*\phi\rightarrow \psi\rightarrow l\phi^*$ they
obtain the amplitudes for 
$\tilde\psi\rightarrow l \phi^*,l^*\phi$ in the large mass difference limit%
\footnote{%
In section 9 of ref.\cite{pilaftsis1} a similar factorisation of two
particle scattering amplitudes was done (in a different context) to obtain
the amplitude for decays of unrotated fields.}.
In the limit of a large mass difference,
$|M_1^\prime-M_2^\prime|\gg |H_{12}|$, 
the matrix relating
$N_{aR}$ and $\psi_{aR}$ fields 
for ref.\cite{BP} is given
by
\begin{equation}
 \left( \begin{array}{cccccc}
                  1 & -{2 M_2 \widetilde H_{12}\over M_1^2-M_2^2} \\
                  {2 M_1 \widetilde H_{12}\over M_1^2-M_2^2} & 1  \\
              \end{array}    \right)
\label{eq:BPwk}\end{equation}
Though the matrix above differs from that for ref.\cite{FPSW} 
the 
ultimate asymmetry is the same as that of ref.\cite{FPSW} because the
factor of 2 in the off-diagonal entries above compensates for the presence of
the terms proportional to $H_{12}$ in eq.~(\ref{eq:VUS4wk})
\footnote{%
Note that the lagrangian of ref.\cite{FPSW} as given 
in eq.~(\ref{eq:lag}) is twice that of ref.\cite{BP}.}.
We have also obtained the matrix relating
$N_{aL,R}$ and $\psi_{aL,R}$ fields 
in the limit of a small mass difference, 
$|M_1^\prime -M_2^\prime| \ll |H_{12}|$, by diagonalising the projections of
the resummed
propagator as in 
ref.\cite{BP}.  We obtain the same matrix as
refs.\cite{FPSW,SV}, as given in eq.~(\ref{eq:VUSres})%
\footnote{%
In general $S_{RR}$ and $S_{LL}$ do not have the same poles making it
non-trivial to define the particle mass.  We do not address such field
theoretic subtleties.  We only mention that in the limit that 
$|M_1^\prime -M_2^\prime|\ll |H_{12}|$ $S_{RR}$ and $S_{LL}$ do have the same
pole.}.

In ref.\cite{pilaftsis2} the author obtains the resummed propagator
as in ref.\cite{BP} but does not diagonalise it and obtains the 
amplitudes for $N_1\rightarrow l \phi^*, l^* \phi$ by
by truncating the
Green's function in momentum space with
$(S_{11})^{-1}=\pslash-M_1+\hat\Sigma_{11}
-\hat\Sigma_{12}[\pslash-
M_2+\hat\Sigma_{22}]^{-1}\hat\Sigma_{21}$, where the self-energy
corrections to the propagator $\hat\Sigma_{ij}=i \Sigma^{abs}_{ij}$
are defined in eq.~(4.2) of ref.\cite{pilaftsis2}.  In the LSZ formalism
the truncating operator defines the corresponding incoming or outgoing
particle.  This
implies that the incoming neutrino has
a mass that is the (complex) pole of the component 
$S_{11}$ and not of the full resummed
propagator $S$, i.e.,  
one is including only one loop diagonal corrections
to define the neutrino mass and excluding off-diagonal corrections.
Excluding mixing in the incoming state allows one to use the classical 
Boltzmann 
equation as discussed later.
The derivation of the 
LSZ formula in momentum space used in ref.\cite{pilaftsis2} involves integrations over time in intermediate steps, whose range of integration must be modified
from $(-\infty,+\infty)$ to $(0,+\infty)$ if the neutrino fields have a complex
mass%
\footnote{%
Integrating over a semi-infinite time interval has been taken into account
in ref.\cite{CR97} (and, for example, in ref.\cite{mohanty} in a different
context).}.  
However the decay amplitude obtained in ref.\cite{pilaftsis2}
can be embedded in an amplitude involving
only stable particles in asymptotic states
(analogous to the approach in section 9 of ref.\cite{pilaftsis1}), 
if one equates $S_{11}=
1/[\pslash -M_1+\hat\Sigma_{11}
-\hat\Sigma_{12}(\pslash-
M_2+\hat\Sigma_{22})^{-1}\hat\Sigma_{21}]$ with
$u_1 \bar {\tilde u}_1/(...)$ and again $u_1$ and $\bar {\tilde u}_1$
satisfy only a diagonal Dirac 
equation%
\footnote{%
For a field with a complex mass $m$, $u \bar {\tilde u}$,
and not $u \bar u$, equals $\pslash +m$.}. 
(``..." is approximately equal to $i M_1 \Gamma_1$ where $\Gamma_1$
includes width effects close to $N_1$ production.)

We reiterate here the nature of the incoming particle in various approaches.
In ref.\cite{CRV96} the incoming particle is an unmixed particle corresponding
to the original lagrangian of the theory, i.e., 
$N_{1L,R}$ of eq.~(\ref{eq:lag}).
In ref.\cite{CR97} the incoming particle is a particle corresponding to
the mixed field of the one loop effective theory, i.e,
$N_{1L,R}$ of eq.~(\ref{eq:Meff}).  In ref.\cite{FPS,FPSW,ma,SV} the 
incoming particle is a particle corresponding to the rotated field
of the one loop effective theory, i.e., $\psi_1$ of eq.~(\ref{eq:MD}) or
its equivalent.  In ref.\cite{BP} the incoming particle corresponds to the
rotated field obtained by diagonalising projections of 
the resummed propagator.  The incoming
particle has a one loop effective mass.  In ref.\cite{pilaftsis2} the incoming
particle is unrotated but has a one loop (complex)
effective mass corresponding to the pole of the $S_{11}$ component
of the resummed propagator.
In the limit that mixing is small
compared to the mass difference, i.e, $|H_{12}|\ll |M_1^\prime-M_2^\prime|$ 
the amplitudes
for decay in
refs.\cite{CRV96,BP,pilaftsis2} are similar though the spinor $u$ in the 
amplitude satisfies $u\bar u=\pslash+m$, or $u\bar {\tilde u}=\pslash+m$, 
with a different $m$ for each
approach.  As implied earlier the amplitude for ref.\cite{FPSW} differs in
form from the others but refs.\cite{CRV96,FPSW,BP,pilaftsis2} give
approximately the same asymmetry in this limit%
\footnote{%
When the incoming particle has a complex mass
there are complications in obtaining the decay rate by squaring the
decay amplitude because
$u\bar u\ne\pslash+m$.  However this does not affect the
asymmetry which involves a ratio of decay 
rates.}.

In the opposite 
limit that $|M_1^\prime-M_2^\prime|\ll|H_{12}|$ the rotating matrices of 
refs.\cite{FPSW,SV,BP} are equivalent
and give the same asymmetry.  It was first shown in ref.\cite{FPSW}
that there is an enhancement in the asymmetry when 
$|M_1-M_2|\le|H_{12}|$ and $|H_{11}-H_{22}|\ll |H_{12}+\tilde H_{12}|$.
We have reevaluated
the asymmetry in this limit and obtained
\begin{equation}
\epsilon_{\psi_1}=-{Im(h_{\alpha 1}^* h_{\alpha 2}h_{\beta 1}^* h_{\beta 2})
\over8\pi}{\eta 
\over \eta^2 [2 |h_{\delta2}|^2 + Re(h_{\gamma
1}^* h_{\gamma 2})]+ 4(g_{\alpha ij}^{abs})^2 Re^2(h_{\epsilon 1}^* h_{\epsilon 2}) |h_{\delta1}+h_{\delta2}|^2 
}
\label{eq:e1}\end{equation}
where $\eta=(M_2-M_1)/M_1$ and the sum over $\alpha,\beta,\gamma,\delta$ and $\epsilon$ is implied%
\footnote {%
This result differs from that of ref.\cite{FPSW} but we have been informed by
Dr. U. Sarkar that there is a correction to the 
calculaton of $\epsilon_{\psi_1}$ in
ref.\cite{FPSW} in this limit.}.  
The expression for the asymmetry
obtained in ref.\cite{pilaftsis2} in the limit $|M_1^2-M_2^2|\ll M_1^2$ is
\begin{equation}
\epsilon_{N_1}=
{Im(h_{\alpha 1}^* h_{\alpha 2}h_{\beta 1}^* h_{\beta 2})
\over 8\pi |h_{\delta 1}|^2} {r_N\over r_N^2 +4 A_{22}^2}
\label{eq:e2}\end{equation}
where $r_N=(M_1^2-M_2^2)/(M_1 M_2)\approx-2\eta$ and 
$A_{ij}=\sum_\alpha h_{\alpha i} h_{\alpha j}^*/(16\pi)$.
In obtaining the asymmetry above, 
$\pslash$ in the invariant amplitude was substituted by $M_1$.
To be consistent with the truncation of the Greens's function
$\pslash$ 
should be replaced by the pole of $S_{11}$.  It is not trivial
to solve for the pole. 
Using a naive 
substitution of $\pslash$ by
$M_1-\hat\Sigma_{11}
+\hat\Sigma_{12}[M_1-
M_2+\hat\Sigma_{22}]^{-1}\hat\Sigma_{21}$, where $\hat\Sigma_{ij}$ are evaluated at
$\pslash=M_1$ and
ignoring terms of order $h_{\alpha i}^4$ at the amplitude level, as in ref.\cite{pilaftsis2}, one obtains%
\footnote{%
Finite temperature effects, which we have not included, can induce
changes in particle
masses which can be relevant \cite{pilaftsis3}.} 
\begin{equation}
\epsilon_{N_1}^\prime=
{Im(h_{\alpha 1}^* h_{\alpha 2}h_{\beta 1}^* h_{\beta 2})
\over 8\pi |h_{\delta 1}|^2} {r_N(1-a)\over r_N^2(1-a)^2 +4(A_{22}-A_{11}+aA_{22})^2}
\, ,\label{eq:e3}\end{equation}
where $a=|A_{12}|^2M_1^2/[(M_1-M_2)^2+A_{22}^2M_1^2]$. 
In the case of hierarchical Yukawa couplings the above result reduces to
the asymmetry in eq.~(\ref{eq:e2}).
In the limit that $A_{11}\sim A_{22}$, or $|h_{\alpha 1}|\sim |h_{\alpha2}|$, and $\eta\sim A_{22}$,
which is comparable to the limit in which the enhancement was seen in
ref.\cite{FPSW}, $a\sim1/2$.  Then the asymmetry reduces to 
\begin{equation}
\epsilon_{N_1}^\prime=
{Im(h_{\alpha 1}^* h_{\alpha 2}h_{\beta 1}^* h_{\beta 2})
\over 8\pi |h_{\delta 1}|^2} {2r_N\over r_N^2 +4 A_{22}^2}\,.
\label{eq:e4}\end{equation}
One can see that the asymmetry in eq.~(\ref{eq:e1}) 
is different from that in eq.~(\ref{eq:e4}).  
This is not surprising as  eq.~(\ref{eq:e1}) represents
the asymmetry in the decay of a $\psi$ particle while
eq.~(\ref{eq:e4}) reflects the asymmetry in the decay of a
particle represented by an unrotated field with a complex mass.  
In the limit of large mixing these need not be equivalent.       

We have also obtained the asymmetry in the small mass difference limit
for the model involving sneutrino decays in ref.\cite{CR97}.  In the limit
that $|M_1^2-M_2^2|\ll M_1^2$ and 
$|\Gamma^2_{11}-\Gamma^2_{22}|\ll |\Gamma^2_{12}|$ (in the notation of 
ref.\cite{CR97}), we find 
\begin{equation}
\epsilon_{N_1}=
{Im(h_{\alpha 1}^* h_{\alpha 2}h_{\beta 1}^* h_{\beta 2})
\over \pi (|h_{\delta 1}|^2+|h_{\delta 2}|^2)} {r_N\over r_N^2 +16[(A_{22}-A_{11})^2+4Im^2 A_{12}]}
\, ,\label{eq:e5}\end{equation}
where $A_{ij}$ and $r_N$ are defined as above.  While comparing this result with the 
asymmetry obtained in other works one must keep in mind that eq.~(\ref{eq:e5})
is the asymmetry obtained from sneutrino decays and that the model in 
ref.\cite{CR97} has twice as many decay channels for the sneutrinos as compared
to the neutrinos in other works.

We would also like to add a note of caution when one uses 
the effective Hamiltonian approach and the Boltzmann equation
to obtain the lepton asymmetry.  
The Boltzmann equation which is used
to calculate the number densities of various species assumes that the nature of
the species does not change over time.  The l.h.s. of the Boltzmann equation is the
time derivative of the number density of a species 
and one  has to be careful as to exactly what
one means by $n_N(t)$ or $n_\psi(t)$ at times when $N_a$ or $\psi_a$ may not
be the physical states.  
This comment applies to scenarios where one starts with $N$ states 
at
$t_{nr}$ which evolve subsequently as $\psi$ states, as well as to
ref.\cite{FP} where one obtains the asymmetry
by evolving the $\psi$ number density from when
the neutrinos are relativistic. 

It has been argued in ref.\cite{pilaftsis3} that decoherence effects due to
interactions with the thermal universe justify excluding mixing effects
in the incoming neutrino state.  Excluding mixing in the incoming state allows one to
work with a classical (diagonal) Boltzmann equation.

\section{Summary}

In conclusion, we have argued that in leptogenesis scenarios where the heavy 
neutrinos or scalars are relativistic when they go out of equilibrium one should
calculate the asymmetry in the decay of the lightest unrotated species.  We 
have pointed out certain issues related to the definition of the decay amplitude
for eigenkets of the non-Hermitian effective Hamiltonian and the existence
of a Fock space for mixed fields.  As different authors have used various 
approaches to rotate the mixed fields we have compared the rotating matrices of
refs.\cite{FPSW,SV,BP} and discussed their similarities and differences. 
Finally we have re-evaluated the asymmetry in the small mass difference limit
using different approaches.

\noindent {\bf Acknowledgement} \\ RR would like to thank
W. Buchmuller, L. Covi, M. Flanz, A. Pilaftsis, M. Plumacher,
S. Mohanty, S. Rindani, U. Sarkar and R. Vaidya
for clarifying and helpful discussions.



\begin{thebibliography}{99}

\bibitem{reviews} E. W. Kolb and M. S. Turner, Ann. Rev. Nucl. Part. Sci.
{\bf 33} (1983) 645; A. D. Dolgov, Phys. Rep. {\bf 222} (1992) 309;
V. A. Rubakov and M. E. Shaposhnikov, Usp. Fiz. Nauk {\bf 166} (1996) 493,
Phys. Usp. {\bf 39} (1996) 461; A. Riotto, hep-ph/9807454.

\bibitem{leptov} M. Fukugita and T. Yanagida, Phys. Lett. {\bf B 174}
  (1986) 45;
P.  Langacker, R.D.  Peccei and T.  Yanagida, Mod.
  Phys.  Lett.  {\bf  A  1}  (1986)  541;  M.A.  Luty, Phys.
  Rev.  {\bf D 45}  (1992)  455;  C. Vayonakis, Phys. Lett. {\bf B286} (1992) 92; R.N.  Mohapatra  and X.  Zhang,
  Phys.  Rev.  {\bf D 46} (1992) 5331

\bibitem{IKS} One loop self-energy corrections were included in baryogenesis
models in A. Yu. Ignat'ev, V. A. Kuz'min and M. E. Shaposhnikov, 
JETP Lett. {\bf 30} (1979) 688 and in 
F. J. Botella and J. Roldan, Phys. Rev. {\bf D 44} (1991) 966.
Asymmetry due to mixing is also mentioned
in section 2.4.4 of E. W. Kolb and S. Wolfram, Nucl. Phys. {\bf B 172}
(1980) 224.  Self-energy corrections were first included in leptogenesis
models in ref.\cite{LS}.

\bibitem{LS} J. Liu and G. Segre, Phys. Rev. {\bf D 48} (1993) 4609.

\bibitem{CRV96}  L. Covi, E. Roulet and F. Vissani, Phys. Lett.
  {\bf B 384}, 169 (1996).

\bibitem{FPS} M. Flanz, E.A. Paschos and U. Sarkar, Phys. Lett.
  {\bf B 345} (1995) 248; Phys. Lett. {\bf B 384} (1996) 487 (erratum).

\bibitem{FPSW} M. Flanz, E.A. Paschos, U.  Sarkar
  and J. Weiss, Phys. Lett. {\bf B 389}, 693 (1996).

\bibitem{ma} E. Ma and U. Sarkar, Phys. Rev. Lett. {\bf 80} (1998) 5716.

\bibitem{SV} U. Sarkar and R. Vaidya, Phys. Lett. {\bf B 442} (1998) 243.

\bibitem{CR97}  L. Covi and E. Roulet, Phys. Lett. {\bf B 399} (1997) 113.

\bibitem{KRS} The possibility of such an ehancement  was first mentioned in
V. A. Kuzmin, V. A. Rubakov and M. E. Shaposhnikov,
Phys. Lett. {\bf B 155} (1985) 36.

\bibitem{BP}  W. Buchm\"{u}ller and M. Pl\"{u}macher, Phys. Lett. {\bf B 431} 
(1998) 354.

\bibitem{pilaftsis2}  A. Pilaftsis, Phys. Rev. {\bf D 56} (1997) 5431.

\bibitem{paschos} E. A. Paschos and U. Turke, Phys. Rept. {\bf 178} (1989) 145.

\bibitem{kniehlpilaftsis} B. A. Kniehl and A. Pilaftsis, Nucl. Phys. 
{\bf B 474} (1996) 268.  

\bibitem{LS2} J. Liu and G. Segre, Phys. Rev. {\bf D 49} (1994) 1342.

\bibitem{enz} C. P. Enz and R. R. Lewis, Helv. Phys. Acta {\bf 38} (1965) 860;
in {\it CP Violation}, edited by L. Wolfenstein (North-Holland, Amsterdam,
1989), p. 58.

\bibitem{kimetal} C. Giunti, C. W. Ki and U. W. Lee, Phys. Rev {\bf D 45}, 
(1992) 2414.

\bibitem{FP}  M. Flanz and E.A. Paschos, Phys. Rev. {\bf D 58} (1998) 113009.

\bibitem{veltman} M. Veltman, Physica {\bf 29} (1963) 186.

\bibitem{pilaftsis1}  A. Pilaftsis, Nucl. Phys. {\bf B 504} (1997) 61.

\bibitem{peskin}  M. E. Peskin and D. V. Schroeder, {\it An Introduction to
Quantum Field Theory} (Addison-Wesley Publishing Company, Reading, 1995).

\bibitem{frereetal} J. M. Frere, F.-S. Ling, M. H. G. Tytgat and
V. Van Elewyck, Phys. Rev. {\bf D 60} (1999) 016005.

\bibitem{mohanty} W. Grimus, P. Stockinger and S. Mohanty,
Phys. Rev. {\bf D59} (1998) 013011.


\bibitem{pilaftsis3} A. Pilaftsis, Int. J. Mod. Phys. {\bf A 14} (1999) 1811.

\end{thebibliography}
\end{document}